\documentstyle[12pt,subeqnarray,psfig]{article}
\begin{document}
%
%
%
%
%
\newenvironment{lefteqnarray}{\arraycolsep=0pt\begin{eqnarray}}
{\end{eqnarray}\protect\aftergroup\ignorespaces}
\newenvironment{lefteqnarray*}{\arraycolsep=0pt\begin{eqnarray*}}
{\end{eqnarray*}\protect\aftergroup\ignorespaces}
\newenvironment{leftsubeqnarray}{\arraycolsep=0pt\begin{subeqnarray}}
{\end{subeqnarray}\protect\aftergroup\ignorespaces}
\newcommand{\displayfrac}[2]{\frac{\displaystyle #1}{\displaystyle #2}}
\newcommand{\diff}{{\rm\,d}}
\newcommand{\img}{{\rm i}}
\newcommand{\appleq}{\stackrel{<}{\sim}}
\newcommand{\appgeq}{\stackrel{>}{\sim}}
\newcommand{\Int}{\mathop{\rm Int}\nolimits}
\newcommand{\Nint}{\mathop{\rm Nint}\nolimits}
\newcommand{\Min}{\mathop{\rm min}\nolimits}
\newcommand{\Max}{\mathop{\rm max}\nolimits}
\newcommand{\Sgn}{\mathop{\rm Sgn}\nolimits}
\newcommand{\arcsinh}{\mathop{\rm arcsinh}\nolimits}
\newcommand{\vers}{\mathop{\overrightarrow{\rm vers}}\nolimits}

\title{R fluids}   

\author{{R. Caimmi}\footnote{
%
%
{\it Dipartimento di Astronomia, Universit\`a di Padova,
              Vicolo Osservatorio 2, I-35122 Padova, Italy -} 
              email: caimmi@pd.astro.it}
\phantom{agga}}

\maketitle
\begin{quotation}
\section*{}
\begin{Large}
\begin{center}

Abstract

\end{center}
\end{Large}
\begin{small}

A theory of collisionless fluids is developed
in a unified picture, where nonrotating
$(\widetilde{\Omega_1}=\widetilde{\Omega_2}=
\widetilde{\Omega_3}=0)$ figures with anisotropic
$(\sigma_{11}=\sigma_{22}=\sigma_{33})$
random velocity component distributions and
rotating
$(\widetilde{\Omega_1}\ne\widetilde{\Omega_2}\ne
\widetilde{\Omega_3})$ figures with isotropic
$(\sigma_{11}\ne\sigma_{22}\ne\sigma_{33})$
random velocity component distributions, make
adjoints configurations to the same system.
R fluids are defined as
ideal, self-gravitating fluids satisfying the
virial theorem assumptions, in presence of systematic
rotation around each principal axis of inertia.
To this aim, mean and rms angular velocities and
mean and rms tangential velocity components are
expressed, by weighting on the moment of
inertia and the mass, respectively.   The figure
rotation is defined as the mean angular velocity,
weighted on the moment of inertia, with respect
to a selected axis.   The generalized tensor virial equations
(Caimmi \& Marmo 2005) are formulated for R fluids
and further attention is devoted to axisymmetric
configurations where, for selected coordinate
axes, a variation in figure rotation has to be
counterbalanced by a variation in anisotropy excess and vice
versa.   A microscopical analysis of systematic
and random motions is performed under a few
general hypotheses, by reversing the sign of
tangential or axial velocity components of an
assigned fraction of particles, leaving the
distribution function and other parameters
unchanged (Meza 2002).   The application of
the reversion process to tangential velocity
components, is found to imply the conversion
of random motion rotation kinetic energy into
systematic motion rotation kinetic energy.
The application of
the reversion process to axial velocity
components, is found to imply the conversion
of random motion translation kinetic energy into
systematic motion translation kinetic energy, and the
loss related to a change of reference frame is
expressed in terms of systematic (imaginary) motion 
rotation kinetic energy.   A number of special
situations are investigated with further detail.
It is found that a R fluid always admits an
adjoint configuration where figure rotation
occurs around only one principal axis of
inertia (R3 fluid),
which implies that all the results related to
R3 fluids (Caimmi 2007) may be extended to R
fluids.   Finally, a procedure is sketched for
deriving the spin parameter distribution
(including imaginary rotation) from
a sample of observed or simulated large-scale
collisionless fluids i.e. galaxies and galaxy
clusters.

\noindent
{\it keywords - galaxies: clusters - galaxies:
haloes - stars: stellar systems.}

\end{small}
\end{quotation}
%

\section{Introduction}\label{intro}
Due to particle shocks, collisional fluids
(e.g., stars, gas clouds) exhibit an isotropic
stress tensor $(\sigma_{11}^2=\sigma_{22}^2=
\sigma_{33}^2)$, where $\sigma_{pp}^2$ is the
rms random velocity component on the axis,
$x_p$.   The absence of particle shocks
(leaving aside extreme situations, such as
high-density galactic nuclei) makes a different
situation in collisionless fluids (e.g.,
galaxies, galaxy clusters), where the stress
tensor is - in general - anisotropic $(\sigma_
{11}^2\ne\sigma_{22}^2\ne\sigma_{33}^2)$.
The shape of the body is determined by systematic
rotation, which is quantified by a spin parameter
(null for nonrotating configurations), and/or by
the difference between stress tensor diagonal
components, or any equivalent anisotropy indicator,
related to the rotation and an equatorial
principal axis of inertia, respectively (null
for configurations where the random velocity
component distribution is isotropic).   A
description of collisionless fluids based
on the equivalence of systematic and random
motions  with respect to the shape, appears
to be highly rewarding and it would provide
further insight on the properties of stellar
and galaxy systems.

In an earlier attempt (Caimmi 1996) the stress
tensor has been expressed as the sum of two
terms, one related to a random (isotropic)
velocity component distribution, and one
other to anisotropic internal motions within
the system.   Further investigation has been
devoted to the simplest situation where the
system is made of two equal components, which
are rotating at the same rate but in opposite
sense.   Then it has been recognized
that the anisotropy excess may be related to
real rotation, if the shape is flattened, and
to imaginary rotation, if the shape is
elongated, with respect to the rotation axis.

A latter approach (Caimmi \& Marmo 2005) has
been restricted to homeoidally striated
density profiles, for which the tensor
virial equations were formulated and
generalized to unrelaxed configurations.
The kinetic-energy tensor has been expressed
as the sum of two terms, one related to
systematic rotation obeying an assigned law,
and one other to the remaining motions e.g.,
random motions, streaming motions, radial
motions.   Finally, an expression of the
spin parameter in terms of the anisotropy
excess, has shown the role of systematic
and remaining motions in flattening or
elongating the shape.

The above mentioned results have been
improved and extended in subsequent
work (Caimmi 2006, hereafter quoted
as C06), where imaginary rotation has
been related to negative anisotropy
excess.   Then sequences of configurations
for which the generalized tensor virial
equations hold, have been determined for
homeoidally striated Jacobi ellipsoids
including prolate shapes induced by
imaginary rotation.   The results from
numerical simulations on the stability
of rapidly rotating spherical configurations
(Meza 2002) have been interpreted in the
light of the theory.   To this respect,
the key argument is that the reversion
(from clockwise to counterclockwise or
vice versa) of tangential velocity
components related to an assigned fraction
of partices, preserves the potential
energy, the kinetic energy, and the
distribution function (Lynden-Bell
1960, 1962; Meza 2002).

The study on homeoidally striated
Jacobi ellipsoids has been extended
to a more general class of bodies
(R3 fluids) in a recent paper (Caimmi
2007, hereafter quoted as C07), where
the contribution of radial and tangential
velocity components on the equatorial
plane was investigated with further
detail.   In addition, mean and rms
(weighted on the moment of inertia)
angular velocity have been defined,
and related to systematic and random
motion tangential kinetic-energy tensor
components, respectively.   Also
for R3 fluids, it has been realized
that the effect of (positive or
negative) anisotropy excess is
equivalent to additional (real or
imaginary) figure rotation.

The current attempt is aimed to
extend the above mentioned results
to a still more general class of
bodies, R fluids, defined as ideal,
self-gravitating, collisionless
fluids where rotation
occurs around each principal axis
of inertia.   It will be found
that R fluids always admit an
adjoint configuration where figure
rotation occurs around a single
principal axis, that is a R3 fluid.
Accordingly, all the results which
hold for R3 fluids may be extended
to R fluids.

The work is organized as follows.
A number of basic definitions are
provided in Sect.\,2, including
the inertia tensor, the angular-velocity
tensor, and the angular-momentum
tensor.   The generalized tensor
virial equations for R fluids are
formulated in Sect.\,3.   The
microscopical analysis of systematic
and random motions , for a collisionless
fluid made of $N$ identical particles,
is performed in Sect.\,4, where a
velocity component reversion process
is defined, and a number of special
situations are analysed in detail
with respect to kinetic energy changes
from random to systematic motions and
vice versa.   A procedure aimed to the
derivation of
the spin parameter distribution
(including imaginary rotation) 
from an assigned sample of observed
or simulated objects, is outlined
in Sect.\,5.   Some concluding
remarks are reported in Sect.\,6.

\section{Angular-velocity and angular-momentum tensor}
\label{vamat}

In the special case of solid bodies, rotation
is rigid and occurs around a single axis which,
in turn, can remain fixed or change its direction.
Accordingly, the angular momentum and the rotation
kinetic energy read (e.g., Landau \& Lifchits 1966,
Chap.\,VI, \S\S\,31-33; hereafter quoted as LL66):
\begin{lefteqnarray}
\label{eq:Jr}
&& J_r=\sum_{s=1}^3I_{rs}^\prime\Omega_s~~;\qquad r=1,2,3~~; \\
\label{eq:Erot}
&& T_{\rm rot}=\frac12\sum_{r=1}^3\sum_{s=1}^3I_{rs}^\prime\Omega_r
\Omega_s~~;
\end{lefteqnarray}
where $\vec{J}=(J_1,J_2,J_3)$ is the
angular-momentum vector, $\overrightarrow{\Omega}=
(\Omega_1,\Omega_2,\Omega_3)$ the
angular-velocity vector, and $I^\prime$
the inertia tensor:
\begin{equation}
\label{eq:Iprs}
I_{rs}^\prime=\int_S\rho(x_1,x_2,x_3)\left[\delta_{rs}
\sum_{r=1}^3x_r^2-x_rx_s\right]\diff^3S~~;
\end{equation}
related to the density profile, $\rho$,
within the volume, $S$, being $\delta_{rs}$
the Kronecker symbol.   The diagonal
components of the inertia tensor,
$I_{11}^\prime$, $I_{22}^\prime$,
$I_{33}^\prime$, are
the moments of inertia with respect
to the axes, $x_1$, $x_2$, $x_3$,
respectively.

In addition, the inertia tensor is
symmetric of second rank, which
implies the existence of a reference
frame, $({\sf O}^\prime\,X_1\,X_2\,X_3)$,
where the inertia tensor is diagonal
(LL66, Chap.\,VI, \S\,32):
\begin{equation}
\label{eq:Ipd}
I_{rs}^\prime=\delta_{rs}I_{rr}^\prime~~;
\end{equation}
the coordinate axes coincide with the
principal axes of inertia, and the
diagonal components define the
principal moments of inertia.
Accordingly, Eqs.\,(\ref{eq:Jr}) and
(\ref{eq:Erot}) reduce to:
\begin{lefteqnarray}
\label{eq:Jd}
&& J_r=\sum_{s=1}^3I_{rr}^\prime\Omega_r~~;\qquad r=1,2,3~~; \\
\label{eq:Erod}
&& T_{\rm rot}=\frac12\left(I_{11}^\prime\Omega_1^2+
I_{22}^\prime\Omega_2^2+I_{33}^\prime\Omega_3^2\right)~~;
\end{lefteqnarray}
where, in addition (LL66, Chap.\,VI, \S\,32):
\begin{equation}
\label{eq:Ipds}
I_{rr}^\prime\le I_{ss}^\prime+I_{tt}^\prime~~;\quad r=1,2,3~~;
\quad s=2,3,1~~;\quad t=3,1,2~~;
\end{equation}
with regard to the body under consideration.

The inertia tensor has been defined
in a different way, as (e.g., Chandrasekhar
1969, Chap.\,2, \S\,9; Binney \& Tremaine
1987, Chap.\,4, \S\,3):
\begin{equation}
\label{eq:Irs}
I_{rs}=\int_S\rho(x_1,x_2,x_3)x_rx_s\diff^3S~~;
\end{equation}
and the combination of Eqs.\,(\ref{eq:Iprs}) and
(\ref{eq:Irs}) yields:
\begin{lefteqnarray}
\label{eq:IpI1}
&& I_{rs}^\prime=\delta_{rs}\sum_{r=1}^3I_{rr}-I_{rs}~~; \\
\label{eq:IpI2}
&& I_{rr}^\prime=I_{ss}+I_{tt}~~;\qquad r\ne s\ne t~~; \\
\label{eq:IpI3}
&& I_{rs}^\prime=-I_{rs}~~;\qquad r\ne s~~;
\end{lefteqnarray}
or:
\begin{lefteqnarray}
\label{eq:IpI4}
&& 2I_{rr}=I_{ss}^\prime+I_{tt}^\prime-I_{rr}^\prime~~;
\qquad r\ne s\ne t~~; \\
\label{eq:IpI5}
&& I_{rs}=-I_{rs}^\prime~~;\qquad r\ne s~~;
\end{lefteqnarray}
which translates one formulation into
the other (e.g., Bett et al. 2007).

In the general case of (collisional or
collisionless) fluids, rotation could
occur different from solid-body, and
around each principal axis of inertia.
Let $({\sf O}\,x_1\,x_2\,x_3)$ be a
generic reference frame and $({\sf O}^
\prime\,X_1\,X_2\,X_3)$ a reference
frame where the origin coincides with
the centre of inertia, and the coordinate
axes coincide with the principal axes
of inertia.    Let the coordinate axes,
$X_1$, $X_2$, $X_3$, be defined as the
principal axes.
Let $\overrightarrow{\Omega_1}$,
$\overrightarrow{\Omega_2}$,
$\overrightarrow{\Omega_3}$, be the
angular-velocity vectors (to be specified
later) related to
the principal axes of inertia.  Let
$\Omega_{rs}$ be the component of
the vector, $\overrightarrow{\Omega_r}$,
on the coordinate axis, $x_s$.   The
$(3\times3)$ tensor, $\Omega_{rs}$,
is defined as the angular-velocity
tensor of the system under consideration,
with respect to the reference frame,
$({\sf O}\,x_1\,x_2\,x_3)$.   Let
$\overrightarrow{\omega_1}$,
$\overrightarrow{\omega_2}$,
$\overrightarrow{\omega_3}$, be the
angular-velocity vectors related to
the coordinate axes, $x_1$, $x_2$,
$x_3$.   The following relation
holds:
\begin{equation}
\label{eq:oms}
\omega_s=\sum_{r=1}^3\Omega_{rs}~~;\qquad s=1,2,3~~;
\end{equation}
and the angular-velocity tensor, $\omega_{rs}=
\Omega_{rs}$, can formally be defined.
Similarly, the $(3\times3)$ angular-momentum
tensor is expressed as:
\begin{lefteqnarray}
\label{eq:Jrs}
&& J_{rs}^\prime=I_{rs}^\prime\omega_{rs}~~; \\
\label{eq:Js}
&& J_s^\prime=\sum_{r=1}^3I_{rs}^\prime\omega_{rs}~~;
\end{lefteqnarray}
where the inertia tensor is related to the
reference frame, $({\sf O}\,x_1\,x_2\,x_3)$.

In the special case where the
reference frame, $({\sf O}\,x_1\,x_2\,x_3)$,
coincides with $({\sf O}^\prime\,X_1\,X_2\,X_3)$,
then $\Omega_{rs}=\delta_{rs}\Omega_{rr}$.
Accordingly,
Eqs.\,(\ref{eq:oms})-(\ref{eq:Js}) reduce to:
\begin{lefteqnarray}
\label{eq:omp}
&& \omega_s=\Omega_s=\Omega_{ss}=\omega_{ss}~~; \\
\label{eq:Jprs}
&& J_{rs}^\prime=\delta_{rs}I_{rr}^\prime\Omega_{rr}~~; \\
\label{eq:Jps}
&& J_s^\prime=J_s=I_{ss}^\prime\Omega_s~~;
\end{lefteqnarray}
where $I_{tt}^\prime=I_{rr}+I_{ss}$,
$r\ne s\ne t$, represents the moment of
inertia with respect to the principal
axis of inertia, $x_t$.   From this
point on, it shall be intended that
the origin coincides with the centre
of inertia, and the coordinate axes
coincide with the principal axes
of inertia.

The rotation kinetic-energy tensor
is defined as:
\begin{equation}
\label{eq:Tpq1}
(T_{\rm rot})_{rs}=\frac12I_{rs}^\prime\Omega_r\Omega_s=
\frac12\delta_{rs}I_{rr}^\prime\Omega_r^2~~;
\end{equation}
where the diagonal components of the
angular-velocity tensor are expressed
as (C07):
\begin{lefteqnarray}
\label{eq:omr}
&& \Omega_r=\widetilde{\Omega_r}=\frac1{I_{rr}^\prime}\int_S
\left\vert\overrightarrow{\Omega_r}(x_1,x_2,x_3,t)\right\vert w_r^2\rho
(x_1,x_2,x_3,t)\diff^3S;~~ r\ne s\ne t;\qquad \\
\label{eq:momr}
&& \left\vert\overrightarrow{\Omega_r}(x_1,x_2,x_3,t)\right\vert=\frac
{v_{\phi_r}(x_1,x_2,x_3,t)}{w_r}~~; \\
\label{eq:wr}
&& w_r=(x_s^2+x_t^2)^{1/2}~~;
\end{lefteqnarray}
and $\Omega_r(x_1,x_2,x_3,t)$
is the mean value related to all the particles
at the time, $t$, within the infinitesimal
volume element, $\diff^3S=\diff x_1\diff x_2
\diff x_3$, centred on the point, ${\sf P}(x_1,
x_2,x_3)$, $v_{\phi_r}$ is the tangential
velocity component on the $({\sf O}\,x_s\,x_t)$
principal plane, and the moment of inertia,
$I_{rr}^\prime=I_{ss}+I_{tt}$, reads:
\begin{equation}
\label{eq:Iprr}
I_{rr}^\prime=\int_Sw_r^2\rho(x_1,x_2,x_3,t)\diff^3S~~;
\qquad r\ne s\ne t~~;
\end{equation}
as expected from the theorem of the mean,
in connection with Eq.\,(\ref{eq:omr}).

Similarly, the mean square diagonal components
of the angular-velocity tensor are expressed as:
\begin{lefteqnarray}
\label{eq:omr2}
&& \Omega_r^2=\widetilde{(\Omega_r^2)}=\frac1{I_{rr}^\prime}\int_S
\left[\overrightarrow{\Omega_r}(x_1,x_2,x_3,t)\right]^2w_r^2\rho(x_1,x_2,x_3,t)
\diff^3S;~r\ne s\ne t;\qquad
\end{lefteqnarray}
and the related variance reads:
\begin{equation}
\label{eq:varo}
\left(\sigma_{\widetilde{\Omega_r}\widetilde{\Omega_r}}\right)^2=
\widetilde{(\Omega_r^2)}-(\widetilde{\Omega_r})^2~~;
\end{equation}
as known from statistics.

At this stage, it may be useful to extend
and generalize the definition of figure
rotation.
\begin{trivlist}
\item[\hspace\labelsep{\bf Figure rotation.}] \sl
Given a R fluid, the figure rotation is defined
as the mean angular velocity, weighted on the
moment of inertia, with respect to a selected
principal axis.
\end{trivlist}

In terms of tangential velocity components,
the counterparts of Eqs.\,(\ref{eq:omr}),
(\ref{eq:omr2}), and (\ref{eq:varo}) read:
\begin{lefteqnarray}
\label{eq:vpm}
&& \widetilde{v_{\phi_r}}=\frac1M\int_S
\left\vert\overrightarrow{\Omega_r}(x_1,x_2,x_3,t)\right\vert w_r\rho
(x_1,x_2,x_3,t)\diff^3S~~; \quad r\ne s\ne t~~; \\
\label{eq:vp2m}
&& \widetilde{(v_{\phi_r}^2)}=\frac1M\int_S
\left[\overrightarrow{\Omega_r}(x_1,x_2,x_3,t)\right]^2w_r^2\rho
(x_1,x_2,x_3,t)\diff^3S;~~r\ne s\ne t;\qquad \\
\label{eq:vav2}
&& \left(\sigma_{\widetilde{v_{\phi_r}}\widetilde{v_{\phi_r}}}\right)^2=
\widetilde{(v_{\phi_r}^2)}-(\widetilde{v_{\phi_r}})^2~~;
\end{lefteqnarray}
and the combination of Eqs.\,(\ref{eq:omr2})
and (\ref{eq:vp2m}); (\ref{eq:omr}) and
(\ref{eq:vpm}); (\ref{eq:varo}) and
(\ref{eq:vav2}); yields:
\begin{leftsubeqnarray}
\slabel{eq:MIa}
&& M\widetilde{(v_{\phi_r}^2)}=I_{rr}^\prime\widetilde{(\Omega_r^2)}~~; \\
\slabel{eq:MIb}
&& M(\widetilde{v_{\phi_r}})^2=I_{rr}^\prime(\widetilde{\Omega_r})^2~~; \\
\slabel{eq:MIc}
&& M\left(\sigma_{\widetilde{\phi_r}\widetilde{\phi_r}}\right)^2=
I_{rr}^\prime\left(\sigma_{\widetilde{\Omega_r}\widetilde{\Omega_r}}\right)^
2~~;
\label{seq:MI}
\end{leftsubeqnarray}
which relate tangential velocity components
on the $({\sf O}\,x_s\,x_t)$ principal plane,
to angular velocity components on the $x_r$
principal axis.

\section{The generalized tensor virial equations
for R fluids} \label{gtvt}

Let R fluids be defined as (collisional or
collisionless) ideal self-gravitating fluids
where figure rotation occurs around all
the three principal axes of inertia.   Let
$({\sf O}\,x_1\,x_2\,x_3)$ be a reference
frame where the origin coincides with the
centre of inertia, and the coordinate axes
coincide with the principal axes of inertia.
Then the mean radial velocity components
must necessarily equal zero:
\begin{equation}
\label{eq:vwst}
\overline{v_{w_r}}=0~~;\qquad\overline{(v_{w_r}^2)}=
\left(\sigma_{w_rw_r}\right)^2~~;
\end{equation}
where $v_{w_r}$ is the radial velocity component
on the $({\sf O}\,x_s\,x_t)$ principal plane,
perpendicular to the $x_r$ principal axis.
Let positive and negative radial velocity
components be defined as directed outwards
and inwards, respectively.   The same holds
for the mean tangential velocity components:
\begin{equation}
\label{eq:vpst}
\overline{v_{\phi_r}}=0~~;\qquad\overline{(v_{\phi_r}^2)}=
\left(\sigma_{\phi_r\phi_r}\right)^2~~;
\end{equation}
even in presence of systematic rotation.
Let positive and negative tangential velocity
components be defined as rotating
counterclockwise and clockwise, respectively.

The kinetic-energy tensor may be expressed
as the sum of two contributions: one, related
to systematic motions, and one other, related
to random motions (C07).   The result is:
\begin{equation}
\label{eq:Tkk}
T_{k_sk_t}=(T_{\rm sys})_{k_sk_t}+(T_{\rm rdm})_{k_sk_t}~~;
\qquad k=w,\phi~~;
\end{equation}
where the terms on the right-hand side,
using Eqs.\,(\ref{seq:MI}) and (\ref
{eq:vwst}), can be expressed as:
\begin{leftsubeqnarray}
\slabel{eq:Trwa}
&& (T_{\rm sys})_{w_sw_t}=0~~; \\
\slabel{eq:Trwb}
&& (T_{\rm rdm})_{w_sw_t}=\frac12\delta_{st}M\left(\sigma_{w_sw_s}\right)^2~~;
\label{seq:Trw}
\end{leftsubeqnarray}
\begin{leftsubeqnarray}
\slabel{eq:Tspa}
&& (T_{\rm sys})_{\phi_s\phi_t}=\frac12\delta_{st}I_{ss}^\prime
(\widetilde{\Omega_{s}})^2~~; \\
\slabel{eq:Tspb}
&& (T_{\rm rdm})_{\phi_s\phi_t}=\frac12\delta_{st}I_{ss}^\prime
\left(\sigma_{\widetilde{\Omega_{s}}\widetilde{\Omega_{s}}}\right)^2~~;
\label{seq:Tsp}
\end{leftsubeqnarray}
keeping in mind that nondiagonal components
are null in the case under discussion, only
diagonal components shall be considered from
this point on.   The combination of
Eqs.\,(\ref{eq:varo}) and (\ref{seq:Tsp})
yields:
\begin{equation}
\label{eq:Tpr}
T_{\phi_r\phi_r}=\frac12I_{rr}^\prime
\widetilde{(\Omega_{r}^2)}~~;
\end{equation}
which depends on the density profile via
the moment of inertia, $I_{rr}^\prime$,
and the tangential velocity component 
distribution via the mean square angular
velocity, $\widetilde{(\Omega_{r}^2)}$,
regardless from the fraction of systematic
and random motions.

In terms of the contributions related to
the axial components of the kinetic-energy
tensor, $T_{ss}$ and $T_{tt}$, Eqs.\,(\ref
{eq:Tkk}), (\ref{seq:Tsp}), and (\ref
{eq:Tpr}) read:
\begin{leftsubeqnarray}
\slabel{eq:Tpla}
&& (T_{\phi_r\phi_r})_{\ell\ell}=\frac12I_{\ell\ell}
\widetilde{(\Omega_{r}^2)}~~;\qquad\ell=s,t~~; \\
\slabel{eq:Tplb}
&& [(T_{\rm sys})_{\phi_r\phi_r}]_{\ell\ell}=\frac12I_{\ell\ell}
\widetilde{(\Omega_r)}^2~~;\qquad\ell=s,t~~; \\
\slabel{eq:Tplc}
&& [(T_{\rm rdm})_{\phi_r\phi_r}]_{\ell\ell}=\frac12I_{\ell\ell}\left[
\widetilde{(\Omega_{r}^2)}-(\widetilde{\Omega_{r}})^2
\right]~~;\qquad\ell=s,t~~;
\label{seq:Tpl}
\end{leftsubeqnarray}
where Eq.\,(\ref{eq:IpI2}) has been used.

The invariance of a vector with respect
to a change of the reference frame,
implies the validity of the relations (C07):
\begin{lefteqnarray}
\label{eq:vst1}
&& \overline{(v_{w_r}^2)}+\overline{(v_{\phi_r}^2)}=
\overline{(v_s^2)}+\overline{(v_t^2)}~~; \\
\label{eq:vst2}
&& (\overline{v_{w_r}})^2+(\overline{v_{\phi_r}})^2=
(\overline{v_s})^2+(\overline{v_t})^2~~; \\
\label{eq:vast}
&& \left(\sigma_{w_rw_r}\right)^2+\left(\sigma_{\phi_r\phi_r}\right)^2=
\left(\sigma_{ss}\right)^2+\left(\sigma_{tt}\right)^2~~;
\end{lefteqnarray}
where the velocity components on the
$x_s$ and $x_t$ principal axes are
labelled by the indices, $s$ and $t$,
respectively.

The combination of Eqs.\,(\ref{eq:varo}),
(\ref{eq:vav2}), and (\ref{eq:vst1})-(\ref
{eq:vast}) yields:
\begin{equation}
\label{eq:vawr}
\left(\sigma_{w_rw_r}\right)^2=\left(\sigma_{ss}\right)^2+\left(\sigma_{tt}
\right)^2-\frac{I_{rr}^\prime}M\left[\widetilde{(\Omega_r^2)}-
(\widetilde{\Omega_r})^2\right]~~;
\end{equation}
which makes Eqs.\,(\ref{eq:Tkk}) and
(\ref{seq:Trw}) translate into:
\begin{leftsubeqnarray}
\slabel{eq:Twwa}
&& (T_{w_rw_r})_{\ell\ell}=[(T_{\rm rdm})_{w_rw_r}]_{\ell\ell}=
\frac12M\sigma_{\ell\ell}^2-\frac12I_{\ell\ell}\left[
\widetilde{(\Omega_{r}^2)}-(\widetilde{\Omega_r})^2\right]~;
~~\ell=s,t~;\qquad \\
\slabel{eq:Twwb}
&& [(T_{\rm sys})_{w_rw_r}]_{\ell\ell}=0~~;\qquad\ell=s,t~~;
\label{seq:Tww}
\end{leftsubeqnarray}
in terms of the contributions related to
the axial components of the kinetic-energy
tensor, $T_{ss}$ and $T_{tt}$.

The generalized tensor virial equations of
the second order can be formulated, extending
the procedure used for R3 fluids (C07).   The
result is:
\begin{lefteqnarray}
\label{eq:Itt}
&& I_{rr}\left[(\widetilde{\Omega_s})^2+(\widetilde{\Omega_t})^2\right]+
M\zeta_{rr}\sigma^2+(E_{\rm pot})_{rr}=0~~; \\   
\label{eq:sig2}
&& \sigma^2=\sigma_{11}^2+\sigma_{22}^2+\sigma_{33}^2~~;
\end{lefteqnarray}
\begin{leftsubeqnarray}
\slabel{eq:zrra}
&& \zeta_{pp}=\frac{(\widetilde{T}_{\rm rdm})_{pp}}{T_{\rm rdm}}=
\frac{\sigma_{pp}^2}{\sigma^2}~~;\qquad p=1,2,3~~; \\
\slabel{eq:zrrb}
&& \zeta_{11}+\zeta_{22}+\zeta_{33}=\frac{\widetilde{T}_{\rm rdm}}
{T_{\rm rdm}}=\frac{\widetilde{\sigma}^2}{\sigma^2}=\zeta~~;
\label{seq:zrr}
\end{leftsubeqnarray}
where mean angular velocity components on the
$x_r$ principal axis are due to systematic
rotation around $x_s$ and $x_t$ axes, $(E_{\rm
pot})_{rr}$ is the self potential-energy
tensor, $\zeta_{rr}$ may be conceived as
generalized anisotropy parameters (Caimmi
\& Marmo 2005; C06; C07), and $\widetilde
{T}_{\rm rdm}$ is the effective random
kinetic energy i.e. the right amount needed
for an instantaneous configuration to
satisfy the usual tensor virial equations
of the second order, defined by the
effective anisotropy parameters (C07):
\begin{leftsubeqnarray}
\slabel{eq:zerra}
&& \widetilde{\zeta}_{pp}=\frac{(\widetilde{T}_{\rm rdm})_{pp}}
{\widetilde{T}_{\rm rdm}}=
\frac{\zeta_{pp}}{\zeta}~~;\qquad p=1,2,3~~; \\
\slabel{eq:zerrb}
&& \widetilde{\zeta}_{11}+\widetilde{\zeta}_{22}+\widetilde{\zeta}_{33}=1~~;
\label{seq:zerr}
\end{leftsubeqnarray}
and the condition, $\zeta=1$, or $\zeta_{pp}=
\widetilde{\zeta}_{pp}$, $p=1,2,3$, makes
Eqs.\,(\ref{eq:Itt}) reduce to their standard
counterparts.   To get further insight, a
microscopical analysis is needed.

In the special case of axisymmetric
configurations, $I_{11}=I_{22}$,
$(E_{\rm pot})_{11}=(E_{\rm pot})_{22}$,
and the combination of the related tensor
virial equations, expressed by Eq.\,(\ref
{eq:Itt}), yields:
\begin{equation}
\label{eq:IoMs}
I_{pp}\left[(\widetilde{\Omega_q})^2-(\widetilde{\Omega_p})^2\right]=
M\sigma^2(\zeta_{qq}-\zeta_{pp})~~;\quad p=1,2~~;\quad q=2,1~~;  
\end{equation}
where a figure rotation excess,
$[(\widetilde{\Omega_q})^2-(\widetilde{\Omega_p})^2]$,
is counterbalanced by an anisotropy excess,
$(\zeta_{qq}-\zeta_{pp})$; in particular,
a null figure rotation excess implies a null
anisotropy excess and vice versa.
Accordingly, a flattening on the
$({\sf O}\,x_p\,x_r)$ principal plane,
induced by the figure rotation excess,
has to be counterbalanced by an
elongation on the $x_q$ principal
axis, induced by the anisotropy excess,
to yield an axisymmetric configuration
with respect to the $x_r$ principal axis.

\section{Microscopical analysis of systematic
and random motions}
\label{masr}

Given a collisionless R fluid, let $N$
be the total number of particles and $m$
the mean particle mass in absence of mass
segregation i.e. local and global mean
particle mass coincide.   For simplicity,
the equivalent description (C07) involving
$N$ identical particles of mass, $m$,
shall be considered.   With regard to the
$({\sf O}\,x_s\,x_t)$ principal plane,
let $v_{\phi_r}$ be the tangential velocity
component on the above mentioned plane.
It is worth noting (Meza 2002) that the
distribution function is independent of
the sign of $v_{\phi_r}$, and the whole
set of possible configurations are
characterized by an equal amount of
both kinetic and potential energy.
Numerical simulations show that spherical
systems, even if rapidly rotating, are
dynamically stable after reversion of
the tangential velocity component in an
assigned fraction of particles (Meza 2002).

For sake of simplicity, let the initial
configuration be nonrotating $(\overline
{v_{\phi_r}}=0)$
and with isotropic random velocity component
distribution $(\zeta_{11}=\zeta_{22}=\zeta_{33})$.
In the case under discussion of identical
particles, $m^{(i)}=m$, $1\le i\le N$, the
centre of inertia velocity components,
$v_{Cr}$, equal the related arithmetic means:
\begin{equation}
\label{eq:vCr}
v_{Cr}=\displayfrac{\sum_{i=1}^Nm^{(i)}v_r^{(i)}}
{\sum_{i=1}^Nm^{(i)}}=\displayfrac{m\sum_{i=1}^Nv_r^{(i)}}
{Nm}=\overline{v_r}~~;
\end{equation}
and the moments of inertia, $I_{rr}^\prime$,
reduce to:
\begin{equation}
\label{eq:NIpr}
I_{rr}^\prime=\sum_{i=1}^Nm^{(i)}\left[w_r^{(i)}\right]^2=
m\sum_{i=1}^N\left[w_r^{(i)}\right]^2=MR_{Gr}^2~~;
\end{equation}
where $R_{Gr}=[\overline{w_r^2}]^{1/2}$ is
the curl radius with respect to the $x_r$
axis.

The weighted mean, mean square, and rms tangential
velocity components, expressed by Eqs.\,(\ref
{eq:vpm})-(\ref{eq:vav2}), read:
\begin{lefteqnarray}
\label{eq:tvpr}
&& \widetilde{v_{\phi_r}}=\frac1M\sum_{i=1}^Nm^{(i)}v_{\phi_r}^{(i)}=
\frac mM\sum_{i=1}^Nv_{\phi_r}^{(i)}=\overline{v_{\phi_r}}~~; \\
\label{eq:tv2r}
&& \widetilde{(v_{\phi_r}^2)}=\frac1M\sum_{i=1}^Nm^{(i)}\left[v_{\phi_r}^
{(i)}\right]^2=\frac mM\sum_{i=1}^N\left[v_{\phi_r}^{(i)}\right]^2=
\overline{(v_{\phi_r}^2)}~~; \\
\label{eq:vvpr}
&& \left(\sigma_{\widetilde{\phi_r}\widetilde{\phi_r}}\right)^2=
\overline{(v_{\phi_r}^2)}-(\overline{v_{\phi_r}})^2=
\left(\sigma_{\phi_r\phi_r}\right)^2~~;
\end{lefteqnarray}
and Eqs.\,(\ref{seq:MI}) reduce to:
\begin{leftsubeqnarray}
\slabel{eq:vpRga}
&& \overline{(v_{\phi_r}^2)}=R_{Gr}^2\widetilde{(\Omega_r^2)}~~; \\
\slabel{eq:vpRgb}
&& (\overline{v_{\phi_r}})^2=R_{Gr}^2(\widetilde{\Omega_r})^2~~; \\
\slabel{eq:vpRgc}
&& \left(\sigma_{\phi_r\phi_r}\right)^2=R_{Gr}^2
\left(\sigma_{\widetilde{\Omega_r}\widetilde{\Omega_r}}\right)^2~~;
\label{seq:vpRg}
\end{leftsubeqnarray}
which relate weighted angular velocities around
the $x_r$ axis to mean tangential velocity
components on the $({\sf O}\,x_s\,x_t)$ principal
plane.

At this stage, let the tangential velocity
component of a fraction, $n/N$, of particles,
be reversed in equal sense (from clockwise to
counterclockwise or vice versa), according to
the following assumptions.
\begin{description}
\item[\rm{(i)}\hspace{0.2mm}] Both the number,
$n$, of particles where the tangential
velocity component has been reversed, and
the number, $N-n$, of particles which remain
unchanged, are sufficiently large, $1\ll n
\ll N$, $0\le n\le \Int(N/2)$.
\item[\rm{(ii)}~~] The fraction, $n_k/N_k$,
of particles where the tangential velocity
component has been reversed, within a generic
volume element, $S_k$, is independent of the
volume element, $n_k/N_k=n/N$.
\item[\rm{(iii)}] The system is made of
identical particles, $m^{(i)}=m$, $M=mN$.
\item[\rm{(iv)}~] After tangential velocity
components have been reversed in $n_k$
particles within a generic volume element,
$S_k$, on a total of $N_k$, a second set
of $n_k$ particles (among the remaining
$N_k-n_k$) exists, where the tangential
velocity component of any particle equals
its counterpart belonging to the first set.
\end{description}
In the following, the above process shall
be quoted as ``the reversion process''.

Obviously, mean square tangential velocity
components, $\overline{(v_{\phi_r}^2)}$,
are left unchanged by the reversion process.
On the contrary, mean tangential velocity
components after the reversion process read:
\begin{equation}
\label{eq:mvpr}
\overline{v_{\phi_r}}=\frac1N\sum_{i=1}^Nv_{\phi_r}^{(i)}=
\frac1N\left[\sum_{i=1}^{2n}v_{\phi_r}^{(i)}+
\sum_{i=2n+1}^Nv_{\phi_r}^{(i)}\right]~~;
\end{equation}
where the first sum within brackets relates
to particles where the reversion process has
occurred and their counterparts with equal
tangential velocity components, while the
second sum comprises the remaining particles
and necessarily equals the mean tangential
velocity component before the occurrence of
the reversion process, which is null in the
case under discussion.   Accordingly,
Eq.\,(\ref{eq:mvpr}) reduces to:
\begin{leftsubeqnarray}
\slabel{eq:mvp2a}
&& \overline{v_{\phi_r}}=\frac{2n}N(\overline{v_{\phi_r}})_n~~; \\
\slabel{eq:mvp2b}
&& (\overline{v_{\phi_r}})_n=\frac1{2n}\sum_{i=1}^{2n}v_{\phi_r}^{(i)}=
\frac1n\sum_{i=1}^nv_{\phi_r}^{(i)}~~;
\label{seq:mvp2}
\end{leftsubeqnarray}
keeping in mind that the first sum is
performed on couples of particles with
equal tangential velocity components.

The validity of Eqs.\,(\ref{eq:mvpr})
and (\ref{seq:mvp2}) still maintains
if tangential velocity components,
$v_{\phi_r}$, are replaced by axial
velocity components, $v_r$.   The
combination of Eqs.\,(\ref{eq:vCr})
and (\ref{eq:mvp2a}) yields:
\begin{equation}
\label{eq:vCmr}
v_{Cr}=\overline{v_r}=\frac{2n}N(\overline{v_r})_n~~;
\end{equation}
which is the velocity component of
the centre of inertia with respect
to the $x_r$ principal
axis, after the reversion process.

The total kinetic energy is left
unchanged by the reversion process
but, on the other hand, a fraction
of random motion kinetic energy is
turned into systematic motion kinetic
energy.
In the following, the reversion
process shall be discussed with
further details for
a number of different situations.

\subsection{Tangential velocity
component reversion}
\label{rvpp}

Performing the reversion process
on a given fraction of particles,
$n/N$, with respect to tangential
velocity components, implies the
conversion of
random (rotation) motion kinetic
energy into systematic (rotation)
motion kinetic energy, as:
\begin{equation}
\label{eq:DT}
\Delta(T_{\rm rdm})_{\phi_r\phi_r}=-\Delta(T_{\rm sys})_{\phi_r\phi_r}=
-\frac12M(\overline{v_{\phi_r}})^2=-\frac{2n}Nnm[(\overline{v_{\phi_r}})_
n]^2~~;
\end{equation}
where the remaining parameters are
left unchanged.

The occurrence of the reversion
process implies the following
energy changes:
\begin{lefteqnarray}
\label{eq:DTt}
&& T_{\rm rdm}\to T_{\rm rdm}-\frac{2n}Nnm[(\overline{v_{\phi_r}})_n]^2~~; \\
\label{eq:DTp}
&& (T_{\rm rdm})_{\phi_r\phi_r}\to(T_{\rm rdm})_{\phi_r\phi_r}-
\frac{2n}Nnm[(\overline{v_{\phi_r}})_n]^2~~; \\
\label{eq:DTl}
&& (T_{\rm rdm})_{\ell\ell}\to(T_{\rm rdm})_{\ell\ell}-\frac12\frac{2n}Nnm
[(\overline{v_{\phi_r}})_n]^2~~;\qquad\ell=s,t~~; \\
\label{eq:DTst}
&& T_{\rm sys}\to0+\frac{2n}Nnm[(\overline{v_{\phi_r}})_n]^2~~; \\
\label{eq:DTsp}
&& (T_{\rm sys})_{\phi_r\phi_r}\to0+\frac{2n}Nnm[(\overline
{v_{\phi_r}})_n]^2~~; \\
\label{eq:DTspl}
&& (T_{\rm sys})_{\ell\ell}\to0+\frac12
\frac{2n}Nnm(\overline{v_{\phi_r}})_n^2~~;\qquad\ell=s,t~~;
\end{lefteqnarray}
while the contributions from random radial motions
along the equatorial plane, $(T_{\rm rdm})_
{w_rw_r}$, and the rotation axis, $(T_
{\rm rdm})_{rr}$, remain unchanged.

With the system being relaxed, $\zeta=1$,
in the case under discussion, the generalized
and effective anisotropy parameters,
$\zeta_{pp}$ and $\widetilde{\zeta}_{pp}$,
coincide with their counterparts related
to the usual tensor virial equations, and
Eqs.\,(\ref{eq:zrra}) and (\ref{eq:zerra})
take the explicit form (C06):
\begin{leftsubeqnarray}
\slabel{eq:zeqa}
&& \zeta_{\ell\ell}=\displayfrac{(1/3)T_{\rm rdm}-(2n/N)(n/2)m
[(\overline{v_{\phi_r}})_n]^2}{T_{\rm rdm}-(2n/N)nm[(\overline
{v_{\phi_r}})_n]^2}~~;\qquad\ell=s,t~~; \\
\slabel{eq:zeqb}
&& \zeta_{rr}=\displayfrac{(1/3)T_{\rm rdm}}{T_{\rm rdm}-(2n/N)nm
[(\overline{v_{\phi_r}})_n]^2}~~;
\label{seq:zeq}
\end{leftsubeqnarray}
where the special case, $n=0$, relates
to the initial configuration, characterized
by isotropic random velocity component
distributions $(\zeta_{pp}=1/3)$ and no
figure rotation.

In the extreme case where the reversion
process is completed, $n=N/2$, the changes
expressed by Eqs.\,(\ref{eq:DTt})-(\ref
{eq:DTspl}) take the form:
\begin{lefteqnarray}
\label{eq:DNt}
&& T_{\rm rdm}\to T_{\rm rdm}-\frac{N}2m[(\overline{v_{\phi_r}})_{N/2}]^2~~;
\\
\label{eq:DNp}
&& (T_{\rm rdm})_{\phi_r\phi_r}\to(T_{\rm rdm})_{\phi_r\phi_r}-
\frac{N}2m[(\overline{v_{\phi_r}})_{N/2}]^2~~; \\
\label{eq:DNl}
&& (T_{\rm rdm})_{\ell\ell}\to(T_{\rm rdm})_{\ell\ell}-\frac{N}4m
[(\overline{v_{\phi_r}})_{N/2}]^2~~;\qquad\ell=s,t~~; \\
\label{eq:DNst}
&& T_{\rm sys}\to0+\frac{N}2m[(\overline{v_{\phi_r}})_{N/2}]^2~~;
\\
\label{eq:DNsp}
&& (T_{\rm sys})_{\phi_r\phi_r}\to0+\frac{N}2m[(\overline
{v_{\phi_r}})_{N/2}]^2~~; \\
\label{eq:DNspl}
&& (T_{\rm sys})_{\ell\ell}\to0+
\frac{N}4m[(\overline{v_{\phi_r}})_{N/2}]^2~~;\qquad\ell=s,t~~;
\end{lefteqnarray}
similarly, Eqs.\,(\ref{seq:zeq}) take the form:
\begin{leftsubeqnarray}
\slabel{eq:zNqa}
&& \zeta_{\ell\ell}=\displayfrac{(1/3)T_{\rm rdm}-(N/4)m
[(\overline{v_{\phi_r}})_{N/2}]^2}{T_{\rm rdm}-(N/2)m[(\overline
{v_{\phi_r}})_{N/2}]^2}~~;\qquad\ell=s,t~~; \\
\slabel{eq:zNqb}
&& \zeta_{rr}=\displayfrac{(1/3)T_{\rm rdm}}{T_{\rm rdm}-(N/2)m
[(\overline{v_{\phi_r}})_{N/2}]^2}~~;
\label{seq:zNq}
\end{leftsubeqnarray}
in any case, the anisotropy excess,
$\zeta_{\ell\ell}-\zeta_{rr}<0$, is
counterbalanced by figure rotation.

\subsection{Axial velocity component reversion}
\label{rvpa}

Performing the reversion process on a
given fraction of particles, $n/N$,
with respect to axial velocity
components, implies the conversion
of random (translation) motion kinetic
energy into systematic (translation)
motion kinetic energy, as:
\begin{equation}
\label{eq:DTr}
\Delta(T_{\rm rdm})_{rr}=-\Delta(T_{\rm sys})_{rr}=
-\frac12M(\overline{v_r})^2=-\frac{2n}Nnm[(\overline
{v_r})_n]^2~~;
\end{equation}
where the remaining parameters are
left unchanged.

The reversion process implies the
following energy changes:
\begin{lefteqnarray}
\label{eq:DTrt}
&& T_{\rm rdm}\to T_{\rm rdm}-\frac{2n}Nnm[(\overline{v_r})_n]^2~~; \\
\label{eq:DTrp}
&& (T_{\rm rdm})_{rr}\to(T_{\rm rdm})_{rr}-
\frac{2n}Nnm[(\overline{v_r})_n]^2~~; \\
\label{eq:DTrst}
&& T_{\rm sys}\to0+\frac{2n}Nnm[(\overline{v_r})_n]^2~~; \\
\label{eq:DTrsp}
&& (T_{\rm sys})_{rr}\to0+\frac{2n}Nnm[(\overline{v_r})_n]^2~~;
\end{lefteqnarray}
while the contributions from random
motions along the $x_s$ and $x_t$
principal axis, $(T_{\rm rdm})_{ss}$
and $(T_{\rm rdm})_{tt}$, remain
unchanged.

With the system being relaxed, $\zeta=1$,
in the case under discussion, the generalized
and effective anisotropy parameters,
$\zeta_{pp}$ and $\widetilde{\zeta}_{pp}$,
coincide with their counterparts related
to the usual tensor virial equations, and
Eqs.\,(\ref{eq:zrra}) and (\ref{eq:zerra})
take the explicit form:
\begin{leftsubeqnarray}
\slabel{eq:zasa}
&& \zeta_{\ell\ell}=\displayfrac{(1/3)T_{\rm rdm}}{T_{\rm rdm}-(2n/N)nm
[(\overline{v_r})_n]^2}~~;\qquad\ell=s,t~~; \\
\slabel{eq:zasb}
&& \zeta_{rr}=\displayfrac{(1/3)T_{\rm rdm}-(2n/N)nm
[(\overline{v_r})_n]^2}{T_{\rm rdm}-(2n/N)nm[(\overline{v_r})_n]^2}~~;
\label{seq:zas}
\end{leftsubeqnarray}
where the special case, $n=0$, relates
to the initial configuration, characterized
by isotropic random velocity component
distributions $(\zeta_{pp}=1/3)$ and no
figure rotation.

In the extreme case where the reversion
process is completed, $n=N/2$, the changes
expressed by Eqs.\,(\ref{eq:DTrt})-(\ref
{eq:DTrsp}) take the form:
\begin{lefteqnarray}
\label{eq:DNrt}
&& T_{\rm rdm}\to T_{\rm rdm}-\frac{N}2m[(\overline{v_r})_{N/2}]^2~~; \\
\label{eq:DNrp}
&& (T_{\rm rdm})_{rr}\to(T_{\rm rdm})_{rr}-
\frac{N}2m[(\overline{v_r})_{N/2}]^2~~; \\
\label{eq:DNrst}
&& T_{\rm sys}\to0+\frac{N}2m[(\overline{v_r})_{N/2}]^2~~; \\
\label{eq:DNrsp}
&& (T_{\rm sys})_{rr}\to0+\frac{N}2m[(\overline{v_r})_{N/2}]^2~~;
\end{lefteqnarray}
similarly, Eqs.\,(\ref{seq:zas}) take the form:
\begin{leftsubeqnarray}
\slabel{eq:zNaa}
&& \zeta_{\ell\ell}=\displayfrac{(1/3)T_{\rm rdm}}{T_{\rm rdm}-(N/2)m
[(\overline{v_r})_{N/2}]^2}~~;\qquad\ell=s,t~~; \\
\slabel{eq:zNab}
&& \zeta_{rr}=\displayfrac{(1/3)T_{\rm rdm}-(N/2)m
[(\overline{v_r})_{N/2}]^2}{T_{\rm rdm}-(N/2)m[(\overline{v_r})_{N/2}]^2}~~;
\label{seq:zNa}
\end{leftsubeqnarray}
in any case, the anisotropy excess,
$\zeta_{\ell\ell}-\zeta_{rr}>0$, is
counterbalanced by (imaginary) figure
rotation.

\subsection{Change of reference frame
and imaginary rotation}
\label{ivpp}

Performing the reversion process on a
given fraction of particles, $n/N$,
implies the conversion of random motion
(translation) kinetic energy into
systematic motion (translation) kinetic
energy, with respect to the $x_r$
principal axis, according to
Eqs.\,(\ref{eq:DTr})-(\ref{eq:DTrsp}).
The related kinetic-energy tensor
component of the centre of inertia,
by use of Eqs.\,(\ref{eq:vCmr}) and
(\ref{eq:DTr}) reads:
\begin{equation}
\label{eq:TCrr}
(T_C)_{rr}=\frac12M(\overline{v_r})^2=\frac{2n}Nnm[(\overline{v_r})_n]^2=
-\Delta(T_{\rm rdm})_{rr}~~;
\end{equation}
which, in the case under discussion,
coincides with the kinetic energy of
the centre of inertia, $T_C$, being
$(T_C)_{ss}=(T_C)_{tt}=0$.   In the
centre of inertia reference frame
after the reversion process, the
random motion kinetic-energy tensor component
related to the $x_r$ axis, $(T_{\rm
rdm}^\prime)_{rr}$, by use of
Eqs.\,(\ref{eq:vCmr}) and (\ref
{eq:TCrr}), takes the expression:
\begin{equation}
\label{eq:Tprr}
(T_{\rm rdm}^\prime)_{rr}=\frac12m\sum_{i=1}^N\left[v_r^{(i)}-v_{Cr}\right]^2
=(T_{\rm rdm})_{rr}-\frac12M(\overline{v_r})^2=(T_{\rm rdm})_{rr}-(T_C)_{rr}
~~;
\end{equation}
where the kinetic energy, $T_C=
(T_C)_{rr}$, is hidden by the
change of reference frame (e.g.,
LL66, Chap.\,II, \S\,8).

Let the $i$-th particle be at
the distance, $w_r^{(i)}=\{[x_s^
{(i)}]^2+[x_t^{(i)}]^2\}^{1/2}$,
from the $x_r$ principal axis,
with velocity component, $v_r^
{(i)}$.   The imaginary angular
velocity (Caimmi 1996; C06; C07),
${\img}\Omega_r^{(i)}$, can be
defined in such a way the translational
kinetic energy along the $x_r$
axis is counterbalanced by the
imaginary rotational kinetic energy
around the $x_r$ axis, as:
\begin{lefteqnarray}
\label{eq:irot}
&& \frac12m[v_r^{(i)}]^2+\frac12m\left[w_r^{(i)}\right]^2\left[{\img}
\Omega_r^{(i)}\right]^2=0~~; \\
\label{eq:iom}
&& \Omega_r^{(i)}=\frac{v_r^{(i)}}{w_r^{(i)}}~~;
\end{lefteqnarray}
where the index, $i$, is related to the
$i$-th particle, and the factor, i, is
the imaginary unit.   In this view, the
velocity components on the $x_r$ principal
axis may be translated into imaginary
tangential velocity components on the
$({\sf O}\,x_s\,x_t)$ principal plane, as:
\begin{equation}
\label{eq:ivpr}
({\rm i}v_{\phi_r})^2=({\img}w_r\Omega_r)^2=-v_r^2~~;
\end{equation}
according to Eqs.\,(\ref{eq:irot}) and
(\ref{eq:iom}).

Let imaginary rotation around the
$x_r$ axis be imparted to the particles
where the reversion process has been
occurred, and their counterparts with
equal imaginary tangential velocity
components, as prescribed by Eq.\,(\ref
{eq:ivpr}) particularized to the mean
axial velocity component, $\overline
{v_r}$, expressed by Eq.\,(\ref{eq:vCmr}).
The related increment in imaginary kinetic
energy reads:
\begin{equation}
\label{eq:iTsp}
\Delta(T_{\rm sys})_{\phi_r\phi_r}=\frac12M(\overline{\img v_{\phi_r}})^2=
\frac{2n}Nnm[(\overline{\img v_{\phi_r}})_n]^2~~;
\end{equation}
and the combination of Eqs.\,(\ref{eq:DTr}),
(\ref{eq:TCrr}), (\ref{eq:ivpr}), and  
(\ref{eq:iTsp}) yields:
\begin{equation}
\label{eq:iTrp}
\Delta(T_{\rm sys})_{\phi_r\phi_r}=-\Delta(T_{\rm sys})_{rr}=
-(T_C)_{rr}=\Delta(T_{\rm rdm})_{rr}~~;
\end{equation}
the above results may be reduced to a
single statement.
\begin{trivlist}
\item[\hspace\labelsep{\bf Theorem 1.}] \sl
Given a R fluid with isotropic random velocity
distribution and no figure rotation, let
the axial velocity component reversion process
be performed on a given fraction of particles,
$n/N$, with respect to the $x_r$ principal axis.
Then turning to the centre of inertia reference
frame with a kinetic energy loss, $\Delta(T_
{\rm sys})_{rr}=(T_C)_{rr}$, is equivalent to
put the initial configuration into imaginary
rotation around the $x_r$ principal axis,
with square mean tangential velocity component,
$(\overline{\img v_{\phi_r}})^2=-(\overline{v_
r})^2$, with a kinetic energy gain, $\Delta(T_
{\rm sys})_{\phi_r\phi_r}=-(T_C)_{rr}$.
\end{trivlist}
Accordingly, Eqs.\,(\ref{eq:DTrt})-(\ref{eq:DTrsp})
are replaced by the following:
\begin{lefteqnarray}
\label{eq:Tiss}
&& T_{\rm sys}\to0-\frac{2n}Nnm[(\overline{v_{\phi_r}})_n]^2~~; \\
\label{eq:Tisp}
&& (T_{\rm sys})_{\phi_r\phi_r}\to0-\frac{2n}Nnm[(\overline{v_
{\phi_r}})_n]^2~~; \\
\label{eq:Tisl}
&& (T_{\rm sys})_{\ell\ell}\to0-\frac12\frac{2n}Nnm
[(\overline{v_{\phi_r}})_n]^2~~;
\end{lefteqnarray}
while the contributions from random motions
remain unchanged, and the random velocity
distribution remains isotropic $(\zeta_{11}=       
\zeta_{22}=\zeta_{33}=1/3)$.

In the extreme case where the reversion
process is complete, $n=N/2$, and the
maximum amount of available imaginary
rotation has been attained, the changes
expressed by Eq.\,(\ref{eq:Tiss})-(\ref
{eq:Tisl}) take the form:
\begin{lefteqnarray}
\label{eq:TNiss}
&& T_{\rm sys}\to0-\frac N2m[(\overline{v_{\phi_r}})_{N/2}]^2~~; \\
\label{eq:TNip}
&& (T_{\rm sys})_{\phi_r\phi_r}\to0-\frac N2m[(\overline{v_
{\phi_r}})_{N/2}]^2~~; \\
\label{eq:TNil}
&& (T_{\rm sys})_{\ell\ell}\to0-\frac N4m
[(\overline{v_{\phi_r}})_{N/2}]^2~~;
\end{lefteqnarray}
the above results may be reduced to a
single statement.
\begin{trivlist}
\item[\hspace\labelsep{\bf Theorem 2.}] \sl
Given a R fluid with isotropic random velocity
distribution and no figure rotation, let
the axial velocity component reversion process
be performed on a given fraction of particles,
$n/N$, with respect to the $x_r$ principal axis,
and the reference frame changed into
the centre of inertia reference
frame.   Then the resulting configuration with
anisotropy excess, $\zeta_{\ell\ell}-\zeta_{rr}
>0$, Eqs.\,(\ref{seq:zas}), is equivalent to
the initial configuration with null anisotropy
excess, $\zeta_{\ell\ell}-\zeta_{rr}=0$, and
imaginary rotation around the $x_r$ principal
axis, with square mean tangential velocity
component, $(\overline{\img v_{\phi_r}})^2=-
(\overline{v_r})^2$.
\end{trivlist}

\subsection{Anisotropy excess and imaginary
rotation}
\label{rmir}

Given a nonrotating $(\widetilde{\Omega_r}=0)$
isotropic $(\zeta_{11}=\zeta_{22}=\zeta_{33}=
1/3)$ configuration, let the tangential velocity
component on the $({\sf O}\,x_s\,x_t)$
principal plane be reversed in such a way
Eqs.\,(\ref{eq:DNt})-(\ref{seq:zNq}) hold.
Let an equal amount of real and imaginary
tangential velocity component  on the
$({\sf O}\,x_s\,x_t)$ principal plane,
be imparted to each particle for a total
contribution equal to $(2/3)f_rT_{\rm rdm}$
and $-(2/3)f_rT_{\rm rdm}$, respectively,
where $f_r$ is a positive real number, which leaves
the total energy unchanged.   Let the reversion
process be repeated for real tangential velocity
components, to attain a null (real) figure rotation.
The related changes, with respect to the
initial configuration, are:
\begin{lefteqnarray}
\label{eq:fDTr}
&& T_{\rm rdm}\to\left(1+\frac23f_r\right)T_{\rm rdm}~~; \\
\label{eq:fDTp}
&& (T_{\rm rdm})_{\phi_r\phi_r}\to\left(\frac23+\frac23f_r\right)T_{\rm rdm}
~~; \\
\label{eq:fDTl}
&& (T_{\rm rdm})_{\ell\ell}\to\frac12\left(\frac23+\frac23f_r\right)T_
{\rm rdm}~~; \\
\label{eq:fDTsr}
&& T_{\rm sys}\to0-\frac23f_rT_{\rm rdm}~~; \\
\label{eq:fDTs}
&& (T_{\rm sys})_{\phi_r\phi_r}\to0-\frac23f_rT_{\rm rdm}~~; \\
\label{eq:fDTsl}
&& (T_{\rm sys})_{\ell\ell}\to0-\frac12\frac23f_rT_{\rm rdm}~~;
\end{lefteqnarray}
and the related anisotropy parameters
read:
\begin{leftsubeqnarray}
\slabel{eq:zifa}
&& \zeta_{\ell\ell}=\frac{[(1/3)+(1/3)f_r]T_{\rm rdm}}{[1+(2/3)f_r]
T_{\rm rdm}}=\frac{1+f_r}{3+2f_r}~~;\qquad\ell=s,t~~; \\
\slabel{eq:zifb}
&& \zeta_{rr}=\frac{(1/3)T_{\rm rdm}}{[1+(2/3)f_r]
T_{\rm rdm}}=\frac1{3+2f_r}~~; 
\label{seq:zif}
\end{leftsubeqnarray}
where the anisotropy excess, $\zeta_{\ell\ell}-
\zeta_{rr}=f_r/(3+2f_r)>0$, is counterbalanced
by imaginary rotation, according to an initial
configuration with isotropic velocity component
distribution and no figure rotation.   As
application of the above results, two significative
examples shall be taken into consideration.
\begin{trivlist}
\item[\hspace\labelsep{\bf First example.}] \sl
Nonrotating systems flattened
by anisotropic velocity component distribution
$(\sigma_{11}=\sigma_{22}>\sigma_{33})$.
\end{trivlist}
   Let tangential velocity components on the $({\sf O}
\,x_1\,x_2)$ principal plane be reversed in a
convenient fraction of particles, $n/N$, to
yield a convenient amount of figure rotation
together with isotropic velocity component
distribution $(\sigma_{11}^\prime=\sigma_{22}^
\prime=\sigma_{33})$, as sketched in Fig.\,1.
\begin{figure}
\centering
\centerline{\psfig{file=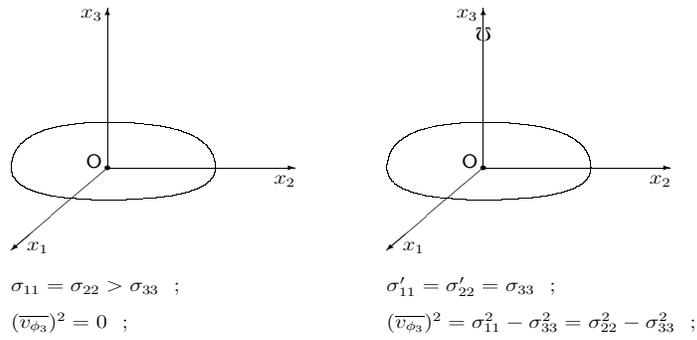,height=180mm,width=150mm}}
\caption{After reversion of tangential
velocity components in a convenient fraction
of particles, with respect to the $({\sf O}x_1x_2)$
principal plane, a configuration flattened
by anisotropic velocity component distribution
$(\sigma_{11}=\sigma_{22}>\sigma_{33})$ with
no figure rotation $(\overline{v_{\phi_3}}=0)$,
left picture, is turned into a configuration
flattened by figure rotation [$\overline{v_
{\phi_3}}=(\sigma_{11}^2-\sigma_{33}^2)^{1/2}=
(\sigma_{11}^2-\sigma_{33}^2)^{1/2}$] with
isotropic velocity component distribution
$(\sigma_{11}^\prime=\sigma_{22}^\prime=
\sigma_{33})$, right picture.   The symbol,
$\mho$, denotes figure rotation around the
related principal axis.}
\label{f:syr1}
\end{figure}

The combination of Eqs.\,(\ref{eq:vvpr}),
(\ref{seq:vpRg}), and (\ref{eq:mvp2a}) yields:
\begin{lefteqnarray}
\label{eq:mORn}
&& (\widetilde{\Omega_r})^2=\frac1{R_{Gr}^2}\frac{4n^2}{N^2}[(\overline
{v_{\phi_r}})_n]^2~~; \\
\label{eq:sORn}
&& \left(\sigma_{\widetilde{\Omega_r}\widetilde{\Omega_r}}\right)^2=
\frac1{R_{Gr}^2}\left(\sigma_{\phi_r\phi_r}\right)^2=
\frac1{R_{Gr}^2}\left\{\overline{(v_{\phi_r}^2)}
-\frac{4n^2}{N^2}[(\overline{v_{\phi_r}})_n]^2\right\}~~;
\end{lefteqnarray}
while the mean square velocity components
are left unchanged by the occurrence of
the reversion process.   The substitution
of Eqs.\,(\ref{eq:mORn}) and (\ref{eq:sORn})
into (\ref{eq:Tplb}) and (\ref{eq:Tplc})
shows the dependence
of systematic and random motion tangential
kinetic-energy tensor components on
the reversion process.

In the case under discussion $(r=3)$,
the random velocity component distribution
has to be isotropic after the reversion
process, which makes Eqs.\,(\ref{eq:DTl})
and (\ref{eq:DTspl}) reduce to:
\begin{lefteqnarray}
\label{eq:Tr3l}
&& (T_{\rm rdm})_{\ell\ell}=\frac M2\left[\sigma_{\ell\ell}^2-\left(
\sigma_{\ell\ell}^2-\sigma_{33}^2\right)\right]=\frac M2\sigma_{33}^2~~;
\qquad\ell=1,2~~; \\
\label{eq:Ts3l}
&& (T_{\rm sys})_{\ell\ell}=\frac M2\left(\sigma_
{\ell\ell}^2-\sigma_{33}^2\right)~~;\qquad\ell=1,2~~; \\
\label{eq:vs3l}
&& \frac{2n}Nnm[\overline{(v_{\phi_r})}_n]^2=\frac12\frac{4n^2}{N^2}M
[(\overline{v_{\phi_r}})_n]^2=\frac M2(\overline{v_{\phi_r}})^2=\frac
M2\left(\sigma_{\ell\ell}^2-\sigma_{33}^2\right);
\end{lefteqnarray}
which defines the configuration after the
reversion process.

\begin{trivlist}
\item[\hspace\labelsep{\bf Second example.}] \sl
Systems elongated by
anisotropic velocity component distribution
$(\sigma_{11}=\sigma_{22}<\sigma_{33})$
with no figure rotation.
\end{trivlist}
Let a convenient
amount of real and imaginary figure rotation,
$\overline{v_{\phi_3}}$ and $\overline{\img
v_{\phi_3}}$, with respect to the $x_3$
principal axis, be imparted to the system.
Then the kinetic energy remains unchanged.
Concerning real rotation, let the reversion
process be performed on one half particles,
leaving imaginary figure rotation together
with isotropic velocity component distribution
$(\sigma_{11}^\prime=\sigma_{22}^
\prime=\sigma_{33})$, as sketched in Fig.\,2.
\begin{figure}
\centering
\centerline{\psfig{file=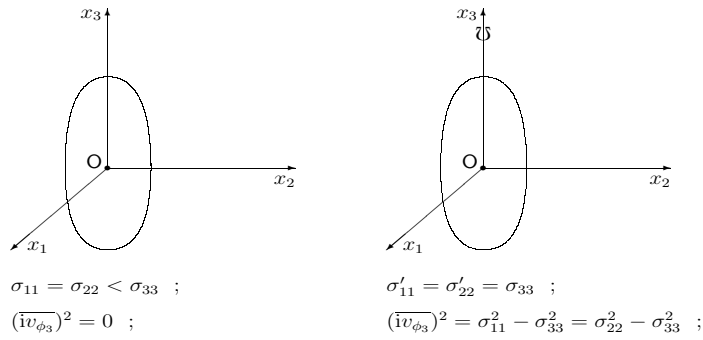,height=180mm,width=150mm}}
\caption{After imparting a convenient amount
of real and imaginary figure rotation,
$\overline{v_{\phi_3}}$ and $\overline{\img
v_{\phi_3}}$, with respect to the $x_3$
principal axis, to the system, and reversing
real tangential velocity components on one
half particles, a configuration elongated
by anisotropic velocity component distribution
$(\sigma_{11}=\sigma_{22}<\sigma_{33})$ with
no figure rotation $(\overline{v_{\phi_3}}=0)$,
left picture, is turned into a configuration
elongated by (imaginary) figure rotation [$\overline
{\img v_{\phi_3}}=(\sigma_{11}^2-\sigma_{33}^2)^{1/2}=
(\sigma_{11}^2-\sigma_{33}^2)^{1/2}$] with
isotropic velocity component distribution
$(\sigma_{11}^\prime=\sigma_{22}^\prime=
\sigma_{33})$, right picture.   The symbol,
$\mho$, denotes figure rotation around the
related principal axis.}
\label{f:syr2}
\end{figure}
Accordingly, Eqs.\,(\ref{eq:mORn}) and
(\ref{eq:sORn}) hold for imaginary
tangential velocity components on the
$({\sf O}\,x_1\,x_2)$ principal plane
and imaginary figure rotation around the $x_3$
principal axis.

In the case under discussion $(r=3)$,
an isotropic velocity component distribution
after the reversion process implies
the validity of Eqs.\,(\ref{eq:Tr3l}),
(\ref{eq:Ts3l}), and (\ref{eq:vs3l}),
where the tangential velocity components
are imaginary $(\sigma_{\ell\ell}<
\sigma_{33})$, and the configuration
after the reversion process is
completely defined.

\subsection{Tangential velocity
component reversion in the general case}
\label{rvgc}

In the general case of anisotropic
velocity component distribution
$(\sigma_{11}\ne\sigma_{22}\ne
\sigma_{33})$ and figure rotation
around each principal axis
$(\widetilde{\Omega_1}\ne\widetilde
{\Omega_2}\ne\widetilde{\Omega_3}\ne0)$,
with regard to the $({\sf O}\,x_s\,x_t)$
principal plane, let the tangential
velocity component reversion process
be applied to one half particles in
such a way no figure rotation around
the $x_r$ principal axis occurs,
$\widetilde{\Omega_r}=0$.   Keeping
in mind Eqs.\,(\ref{eq:tvpr})-(\ref{seq:vpRg}),
the related energy changes read:
\begin{lefteqnarray}
\label{eq:gTrp}
&& (T_{\rm rdm})_{\phi_r\phi_r}\to(T_{\rm rdm})_{\phi_r\phi_r}+\frac12M
(\overline{v_{\phi_r}})^2~~; \\
\label{eq:gTsp}
&& (T_{\rm sys})_{\phi_r\phi_r}\to(T_{\rm sys})_{\phi_r\phi_r}-\frac12M
(\overline{v_{\phi_r}})^2=0~~; \\
\label{eq:vprN}
&& \overline{v_{\phi_r}}=(\overline{v_{\phi_r}})_{N/2}=\frac1N\sum_{i=1}^N
v_{\phi_r}^{(i)}~~; \\
\label{eq:DgTrp}
&& \Delta(T_{\rm sys})_{\phi_r\phi_r}=-\Delta(T_{\rm rdm})_{\phi_r\phi_r}=
-\frac12M(\overline{v_{\phi_r}})^2=-\frac12MR_{Gr}^2(\widetilde
{\Omega_r})^2~~; \\
\label{eq:slg}
&& [(T_{\rm rdm})_{\phi_r\phi_r}]_{\ell\ell}+[(T_{\rm rdm})_{w_rw_r}]_
{\ell\ell}=\frac12M\sigma_{\ell\ell}^2~~;\qquad\ell=1,2~~; \\
\label{eq:srg}
&& (T_{\rm rdm})_{rr}=\frac12M\sigma_{rr}^2~~;
\end{lefteqnarray}
where the rms velocity components,
$\sigma_{\ell\ell}^2$ and $\sigma_{rr}^2$,
are related to the initial configuration.

The changes in anisotropy paramenters,
$\zeta_{pp}=\sigma_{pp}^2/\sigma^2$,
read:
\begin{leftsubeqnarray}
\label{eq:zigl}
&& \zeta_{\ell\ell}\to\frac{\zeta_{\ell\ell}+(1/2)\Delta\zeta_{\ell\ell}}
{1+\Delta\zeta_{\ell\ell}}~~;\qquad\ell=s,t~~; \\
\label{eq:zigr}
&& \zeta_{rr}\to\frac{\zeta_{rr}}{1+\Delta\zeta_{\ell\ell}}~~; \\
\label{eq:Dzig}
&& \Delta\zeta_{\ell\ell}=\frac{(\overline{v_{\phi_r}})^2}{\sigma^2}~~;
\qquad\ell=s,t~~;
\end{leftsubeqnarray}
where the rms velocity, $\sigma^2$, is
related to the initial configuration.

The application of the above procedure
to the $x_1$ and $x_2$ principal axes,
makes the transition from an initial
configuration with rms velocity components,
$\sigma_{11}^2$, $\sigma_{22}^2$, $\sigma_
{33}^2$, and figure rotation around the
principal axes, $\widetilde{\Omega_1}$,
$\widetilde{\Omega_2}$, $\widetilde{\Omega_3}$,
to a final configuration with rms velocity
components, $(\sigma_{11}^\prime)^2$,
$(\sigma_{22}^\prime)^2$, $(\sigma_{33}^
\prime)^2$, and figure rotation around the
principal axes, 0, 0, $\widetilde{\Omega_3}$.
The related energy changes read:
\begin{lefteqnarray}
\label{eq:gTr1}
&& (T_{\rm rdm})_{11}\to(T_{\rm rdm})_{11}+\frac14M
(\overline{v_{\phi_2}})^2~~; \\
\label{eq:gTr2}
&& (T_{\rm rdm})_{22}\to(T_{\rm rdm})_{22}+\frac14M
(\overline{v_{\phi_1}})^2~~; \\
\label{eq:Tr3}
&& (T_{\rm rdm})_{33}\to(T_{\rm rdm})_{33}+\frac14M\left[
(\overline{v_{\phi_1}})^2+(\overline{v_{\phi_2}})^2\right]~~; \\
\label{eq:Tr12}
&& T_{\rm rdm}\to T_{\rm rdm}+\frac12M\left[
(\overline{v_{\phi_1}})^2+(\overline{v_{\phi_2}})^2\right]~~; \\
\label{eq:Trp}
&& (T_{\rm rdm})_{pp}=\frac12M\sigma_{pp}^2~~;\qquad p=1,2,3~~;
\end{lefteqnarray}
and the related changes in anisotropy
parameters are:
\begin{leftsubeqnarray}
\label{eq:zic1}
&& \zeta_{11}\to\frac{\zeta_{11}+(1/2)\Delta\zeta_{11}}
{1+\Delta\zeta_{11}+\Delta\zeta_{22}}~~; \\
\label{eq:zic2}
&& \zeta_{22}\to\frac{\zeta_{22}+(1/2)\Delta\zeta_{22}}
{1+\Delta\zeta_{11}+\Delta\zeta_{22}}~~; \\
\label{eq:Dzic}
\label{eq:zic3}
&& \zeta_{33}\to\frac{\zeta_{33}+(1/2)[\Delta\zeta_{11}+\Delta\zeta_{22}]}
{1+\Delta\zeta_{11}+\Delta\zeta_{22}}~~; \\
&& \Delta\zeta_{11}=\frac{(\overline{v_{\phi_2}})^2}{\sigma^2}~~;\qquad
\Delta\zeta_{22}=\frac{(\overline{v_{\phi_1}})^2}{\sigma^2}~~;~~;
\end{leftsubeqnarray}
the above results may be restricted to a
single statement.
\begin{trivlist}
\item[\hspace\labelsep{\bf Theorem 3.}] \sl
Given a R fluid, a convenient application
of the tangential velocity component reversion
process makes an adjoint configuration where
figure rotation occurs around a single principal
axis, that is a R3 fluid.
\end{trivlist}
Accordingly, the results valid for R3 fluids
(C07) may be extended to the general case of
R fluids.

\section{Discussion}\label{disc}

As suggested in earlier attempts (Caimmi 1996;
C06; C07), the equivalence between a variation in figure
rotation and in anisotropy excess, may provide a
useful tool for the description of collisionless
fluids.   The discussion here shall be focused
on the spin parameter (Peebles 1969, 1971):
\begin{equation}
\label{eq:spin}
\lambda^2=-\frac{J^2E}{G^2M^5}~~;
\end{equation}
where $G$ is the gravitation constant,
$M$ the total mass, $J$ the total angular
momentum, and $E$ the total energy.
The above formulation includes four
possibilities, namely (i) real rotation
$(J^2\ge0)$ and bound system $(E<0)$,
which is the sole currently used in
literature; (ii) imaginary rotation
$(J^2<0)$ and bound system $(E<0)$;
(iii)real rotation $(J^2\ge0)$ and
unbound system $(E\ge0)$; (iv) imaginary
rotation $(J^2<0)$ and unbound system
$(E\ge0)$.   Accordingly, the spin
parameter attains real values in cases
(i) and (iv), and imaginary values
in cases (ii) and (iii).

In the light of the current model,
oblate-like and prolate-like
configurations belong to cases (i)
and (ii) outlined above, while cases
(iii) and (iv) represent unbound
structures for which the virial
equations do not hold.   Then the
comparison with observations and/or
computations, must be restricted to
bound configurations.

The spin parameter distribution is
usually fitted using a lognormal
distribution (e.g., van den Bosh
1998; Gardner 2001; Ballin \&
Steinmetz 2005; Hernandez et al.
2007) or, in general, dependent
on $\log\lambda$ (e.g., Bett et
al. 2007), in dealing with real
rotation.   The inclusion of
imaginary rotation would imply
use of $\lambda^2$ instead of
$\lambda$ as independent variable,
allowing for both positive (real
rotation) and negative (imaginary
rotation) values.

The following procedure should be
followed for calculating $\lambda^2$
from observations and/or computations:
(1) determine the inertia tensor and
the principal axes of inertia for an
assigned matter distribution; (2)
determine the potential-energy tensor;
(3) determine the anisotropy parameters
using the generalized virial equations; (4) perform
the reversion process with respect to
two principal axes of inertia to leave
figure rotation around the third one
(R3 fluid); (5) convert the anisotropy
excess into (real or imaginary) figure
rotation to obtain isotropic velocity
component distribution $(\zeta_{11}=
\zeta_{22}=\zeta_{33})$; (6) evaluate
the spin parameter; (7) act as already
done for all the sample objects; (8)
determine the distribution of the spin
parameter, $P(\lambda^2)$, with respect
to the sample of adjoints configurations,
where the velocity component distribution
is isotropic.

The related results could provide further
insight on the formation and the evolution
of large-scale collisionless fluids, such
as galaxies and galaxy clusters.

\section{Conclusion}\label{conc}

A theory of collisionless fluids has been developed
in a unified picture, where nonrotating
$(\widetilde{\Omega_1}=\widetilde{\Omega_2}=
\widetilde{\Omega_3}=0)$ figures with isotropic
$(\sigma_{11}=\sigma_{22}=\sigma_{33})$
random velocity component distributions and
rotating
$(\widetilde{\Omega_1}\ne\widetilde{\Omega_2}\ne
\widetilde{\Omega_3})$ figures with anisotropic
$(\sigma_{11}\ne\sigma_{22}\ne\sigma_{33})$
random velocity component distributions, make
adjoints configurations to the same system.
R fluids have been defined as
ideal, self-gravitating fluids satisfying the
virial theorem assumptions (e.g., LL66, Chap.\,II,
\S\,10; C07), in presence of figure
rotation around each principal axis of inertia.

To this aim, mean and rms angular velocities and
mean and rms tangential velocity components have
been expressed, by weighting on the moment of
inertia and the mass, respectively.   The figure
rotation has been defined as the mean angular velocity,
weighted on the moment of inertia, with respect
to a selected axis.

The generalized tensor virial equations
(Caimmi \& Marmo 2005) have been formulated for R fluids
and further attention has been devoted to axisymmetric
configurations where, for selected coordinate
axes, a variation in figure rotation has to be
counterbalanced by a variation in anisotropy excess and vice
versa.

A microscopical analysis of systematic
and random motions has been performed under a few
general hypotheses, by reversing the sign of
tangential or axial velocity components of an
assigned fraction of particles, leaving the
distribution function and other parameters
unchanged (Meza 2002).

The application of
the reversion process to tangential velocity
components, has been found to imply the conversion
of random motion rotation kinetic energy into
systematic motion rotation kinetic energy.
The application of
the reversion process to axial velocity
components, has been found to imply the conversion
of random motion translation kinetic energy into
systematic motion translation kinetic energy, and the
loss related to a change of reference frame has been
expressed in terms of systematic motion (imaginary)
rotation kinetic energy.

A number of special
situations have been investigated with further detail.
It has been found that a R fluid always admits an
adjoint configuration where figure rotation occurs
around only one principal axis of inertia (R3 fluid),
which implies that all the results related to
R3 fluids (Caimmi 2007) may be extended to R
fluids.

Finally, a procedure has been sketched for
deriving the spin parameter distribution
(including imaginary rotation) from
a sample of observed or simulated large-scale
collisionless fluids i.e. galaxies and galaxy
clusters.

\end{document}